\documentclass[aps,prd,twocolumn,showpacs,amsmath,amssymb]{revtex4-1}
\usepackage{amsmath}
\usepackage{graphicx}
\usepackage{subfigure}
\usepackage{epstopdf}
\usepackage{color}
\usepackage{multirow}
\usepackage{setspace}
\usepackage{overpic}
\usepackage{amssymb}
\usepackage[bookmarksnumbered, pdfstartview=FitH,colorlinks,urlcolor=blue, citecolor=blue,linkcolor=blue] {hyperref}
\usepackage{lineno}
\usepackage{bm}
\usepackage{rotating}
\usepackage[utf8]{inputenc}
\let\oldequation\equation
\let\oldendequation\endequation

\renewenvironment{equation}
  {\linenomathNonumbers\oldequation}
  {\oldendequation\endlinenomath}

\newcommand{\bes}{BES\uppercase\expandafter{\romannumeral3} }

\begin{document}
%\linenumbers

\title{\bf \boldmath
Measurement of the branching fraction for $\psi(3686)\to \omega K^0_SK^0_S$}

\author{
M.~Ablikim$^{1}$, M.~N.~Achasov$^{10,b}$, P.~Adlarson$^{68}$, S. ~Ahmed$^{14}$, M.~Albrecht$^{4}$, R.~Aliberti$^{28}$, A.~Amoroso$^{67A,67C}$, M.~R.~An$^{32}$, Q.~An$^{64,50}$, X.~H.~Bai$^{58}$
% ...
\\
\vspace{0.2cm}
(BESIII Collaboration)\\
\vspace{0.2cm} {\it
$^{1}$ Institute of High Energy Physics, Beijing 100049, People's Republic of China\\
% ...
$^{4}$ Bochum Ruhr-University, D-44780 Bochum, Germany\\
% ...
$^{10}$ G.I. Budker Institute of Nuclear Physics SB RAS (BINP), Novosibirsk 630090, Russia\\
% ...
$^{14}$ Helmholtz Institute Mainz, Staudinger Weg 18, D-55099 Mainz, Germany\\
% ...
$^{28}$ Johannes Gutenberg University of Mainz, Johann-Joachim-Becher-Weg 45, D-55099 Mainz, Germany\\
% ...
$^{32}$ Liaoning Normal University, Dalian 116029, People's Republic of China\\
% ...
$^{50}$ State Key Laboratory of Particle Detection and Electronics, Beijing 100049, Hefei 230026, People's Republic of China\\
% ...
$^{58}$ University of Jinan, Jinan 250022, People's Republic of China\\
% ...
$^{64}$ University of Science and Technology of China, Hefei 230026, People's Republic of China\\
% ...
$^{67}$ University of Turin and INFN, (A)University of Turin, I-10125, Turin, Italy; (B)University of Eastern Piedmont, I-15121, Alessandria, Italy; (C)INFN, I-10125, Turin, Italy\\
$^{68}$ Uppsala University, Box 516, SE-75120 Uppsala, Sweden\\
% ...
$^{b}$ Also at the Novosibirsk State University, Novosibirsk, 630090, Russia\\
% ...
}
}

\begin{abstract}
 Analyzing $(448.1\pm2.9)\times10^6$ $\psi(3686)$ events collected with the BESIII detector at the BEPCII collider, the $\psi(3686)\to \omega K_{S}^{0}K_{S}^{0}$ decay is observed for the first time. The branching fraction for this decay is determined to be $\mathcal{B}_{\psi(3686)\to \omega K_{S}^{0}K^{0}_{S}}$=$(7.04\pm0.39\pm0.36)$$\times10^{-5}$, where the first uncertainty is statistical and the second is systematic.
\end{abstract}


\maketitle

\oddsidemargin  -0.2cm
\evensidemargin -0.2cm

\section{Introduction}

Experimental studies of charmonium decays are an important tool to study Quantum Chromodynamics (QCD)~\cite{ref::bes3-white-paper}. In perturbative-QCD (pQCD), both $J/\psi$ and $\psi(3686)$ decays to light hadrons mainly proceed through annihilation of the $c\bar{c}$ pair into three gluons, with the decay width proportional to the square of the charmonium wave function at the origin~\cite{ref::aspect3}. One obtains the pQCD 12\% rule for the ratio of branching fractions of $\psi(3686)$ and $J/\psi$ decays to light hadrons:
\begin{equation}
\mathcal{Q}_{h}=\frac{\mathcal{B}_{\psi(3686)\to h}}{\mathcal{B}_{J/\psi\to h}} = \frac{\mathcal{B}_{\psi(3686)\to e^+e^-}}{\mathcal{B}_{J/\psi\to e^+e^-}} = 12.7\%.
\nonumber
\end{equation}
A violation of this rule was first observed by Mark-II in the $J/\psi\to\rho\pi$ decay ~\cite{ref::aspect4} and subsequently found in other decay channels. The inconsistency between pQCD predictions and experimental results is known as the $\rho\pi$ puzzle. Various mechanisms have been proposed to explain the observed discrepancies.
However, no single model provides an explanation of all available experimental results (for a review see Ref.~\cite{ref::aspect5}). % 

Additional measurements of individual branching fractions of $J/\psi$ and $\psi(3686)$ decays provide complementary information to understand the $\rho\pi$ puzzle and further investigate charmonium decay mechanisms~\cite{ref::bes3-white-paper}.
In recent years, a lot of progress has been made in experimental studies of multi-body $J/\psi$ and $\psi(3686)$ decays.
For example, the $J/\psi\to \omega K\bar K$ decays have been studied in Refs.~\cite{Feldman,Falvard} and
the $\psi(3686)\to \omega K^+K^-$ decay has been investigated in Refs.~\cite{Bai,Briere,Ablikim1,bes3_omegaKK}.
In Ref.~\cite{bes3_omegaKK}, $Q_{\omega K^+K^-}$ for the $\psi \to \omega K^+K^-$ decays is determined to be (18.4$\pm$3.7)\%, indicating no significant violation of the 12\% rule.
To date, however, no study of $\psi(3686)\to \omega K^0_SK^0_S$ has been reported.
A first measurement of the branching fraction of $\psi(3686)\to \omega K^0_SK^0_S$ allows to test whether the 12\% rule holds for $\psi\to \omega K^0_SK^0_S$ decays.

The branching fraction of $\psi\to\phi(\omega)K^0\bar K^0$ is expected to be equal to that of $\psi\to \phi(\omega)K^+K^-$ based on isospin symmetry (neglecting possible effects related to continuum production). Earlier measurements summarized in Ref.~\cite{ref::pdg2020} roughly support this relation.
Comparing the obtained branching fraction for $\psi(3686)\to \omega K^0_SK^0_S$ to that for $\psi(3686)\to \omega K^+K^-$ offers an opportunity to search for isospin symmetry violation in $\psi(3686)$ decays.

In this paper, we present the first observation and branching fraction measurement of $\psi(3686)\to \omega K^0_SK^0_S$.
This analysis is carried out  using $(448.1\pm2.9)\times10^6$ $\psi(3686)$ events collected
at the center-of-mass energy of 3.686 GeV with the BESIII detector in 2009 and 2012~\cite{ref::psip-num-inc}.
An inclusive $\psi(3686)$ MC sample, consisting of $506\times10^6$ events, is used to estimate potential backgrounds
from $\psi(3686)$ decays.
The detection efficiency is determined using $5\times10^5$ $\psi(3686)\to\omega K_S^0K_S^0$ decays modeled to reproduce the kinematic distributions found in the data. The $\omega$ decays are described with a Dalitz decay model~\cite{ref::omega}.
~A data sample collected at the center-of-mass energy of 3.65 GeV with an integrated luminosity of 43.88~pb$^{-1}$
is used to estimate the contamination from continuum processes.

\section{BESIII detector and Monte Carlo Simulation}

The BESIII detector~\cite{Ablikim:2009aa} records symmetric $e^+e^-$ collisions
provided by the BEPCII storage ring~\cite{Yu:IPAC2016-TUYA01}, which operates with a peak luminosity of $1\times10^{33}$~cm$^{-2}$s$^{-1}$
in the center-of-mass energy range from 2.0 to 4.95~GeV . BESIII has collected large data samples in this energy region~\cite{ref::bes3-white-paper}. The cylindrical core of the BESIII detector covers 93\% of the full solid angle and consists of a helium-based
 multilayer drift chamber~(MDC), a plastic scintillator time-of-flight
system~(TOF), and a CsI(Tl) electromagnetic calorimeter~(EMC),
which are all enclosed in a superconducting solenoidal magnet
providing a 1.0~T magnetic field. The solenoid is supported by an
octagonal flux-return yoke with resistive plate counter muon
identification modules interleaved with steel. The charged-particle momentum resolution at $1~{\rm GeV}/c$ is
$0.5\%$, and the $dE/dx$ resolution is $6\%$ for electrons
from Bhabha scattering. The EMC measures photon energies with a
resolution of $2.5\%$ ($5\%$) at $1$~GeV in the barrel (end cap)
region. The time resolution in the TOF barrel region is 68~ps, while
that in the end cap region is 110~ps.

Simulated samples produced with the {\sc
geant4}-based~\cite{geant4} Monte Carlo (MC) package, which
includes the geometric description of the BESIII detector and the
detector response, are used to determine the detection efficiency
and to estimate the backgrounds. The simulation includes the beam
energy spread and initial state radiation (ISR) in the $e^+e^-$
annihilations modeled with the generator {\sc
kkmc}~\cite{ref:kkmc}.
The inclusive MC sample consists of the production of the charmonium resonances, and the continuum processes incorporated in {\sc
kkmc}~\cite{ref:kkmc}.
The known decay modes are modeled with {\sc
evtgen}~\cite{ref:evtgen} using branching fractions taken from the
Particle Data Group (PDG)~\cite{ref::pdg2020}, and the remaining unknown decays
from the charmonium states are modeled with {\sc
lundcharm}~\cite{ref:lundcharm}. Final state radiation (FSR)
from charged final-state particles is incorporated with the {\sc
photos} package~\cite{photos}.

\section{Event selection and data analysis}

The two charged pion candidates from the $\omega\to\pi^+\pi^-\pi^0$ decay are required to have a polar angle $\theta$ with respect to the symmetry axis of the detector
of $|\rm{cos\theta}|<0.93$. Furthermore, they are required to originate from the interaction point (IP). The distance of closest approach to the IP must be less than 10\,cm
along the $z$-axis, and less than 1\,cm
in the transverse plane. Particle identification (PID) for charged pion and kaon candidates is performed
using the $dE/dx$ and TOF information.
The confidence levels for pion and kaon hypotheses ($CL_{\pi}$ and $CL_{K}$) are calculated. Pion candidates are required to
satisfy $CL_{\pi}>CL_{K}$.

The $K^0_S$ candidates are reconstructed from two oppositely charged tracks, which are assumed to be pions without imposing PID criteria.
The two charged tracks from the $K_S^0\to\pi^+\pi^-$ decay must satisfy $|\rm{cos\theta}|<0.93$. Their distance of closest approach to the IP along the MDC axis direction is required to be less than 20~cm. A secondary vertex fit is performed for both $K_S^0$ candidates. The chosen $\pi^+\pi^-$ pairs are constrained to originate from a common secondary vertex. Their invariant mass is required to be in the range $(0.486,0.510)~{\rm GeV}/c^2$, which corresponds to an interval of about three times the fitted resolution around the $K^0_S$ nominal mass.
The decay length of the $K^0_S$ candidate is required to be at least two standard deviations of the vertex resolution away from the IP.

The $\pi^0$ candidates are reconstructed from  photon pairs reconstructed from  EMC showers.
Each EMC shower must start within 700~ns of the event start time.
Its energy is required to be greater than 25~MeV in the barrel region  ($|\rm cos\theta|<0.8$)
and 50 MeV in the end cap region ($0.86<|\rm cos\theta|<0.92$).
The opening angle between the photon candidate and the extrapolated impact point of the nearest charged track in the EMC must be greater than $10^{\circ}$.
Each photon pair with an invariant mass within the range $(0.115,\,0.150)$\,GeV$/c^{2}$ is used as a $\pi^0$ candidate.

A five-constraint kinematic fit for the $e^+e^- \to \pi^+\pi^-   \pi^+\pi^-  \pi^+\pi^- \pi^0$ hypothesis is performed including energy-momentum conservation (4C) and with the invariant mass of the two photons constrained to the mass of the $\pi^0$ (1C).
The helix parameters of charged tracks of the MC events are corrected to improve consistency between data and MC simulation (see Ref.~\cite{ref::xiuzheng} for details).
~Events satisfying $\chi^2_{\rm 5C}<80$ are kept for further analysis.
If there are multiple candidates in an event, the one with the smallest  value of  $\chi^2_{\rm 5C}+\chi^2_{K^0_S1}+\chi^2_{K^0_S2}$ is retained,
where $\chi^2_{K^0_S1}$ and $\chi^2_{K^0_S2}$ represent the fit-quality of the secondary vertex fits for the two $K^0_S$ candidates.

Studies of the inclusive $\psi(3686)$ MC sample {\color{blue}using a generic event type analysis tool, TopoAna~\cite{ref::xingyu},} show that those background events surviving all selection criteria contain contributions including an $\omega\to\pi^+\pi^-\pi^0$ decay. These events, however, do not contain a $K^0_SK^0_S$ pair
and can thus be described by $K^0_SK^0_S$ sideband events.

\begin{figure}[htbp]
\centering
\includegraphics[width=1.0\linewidth]{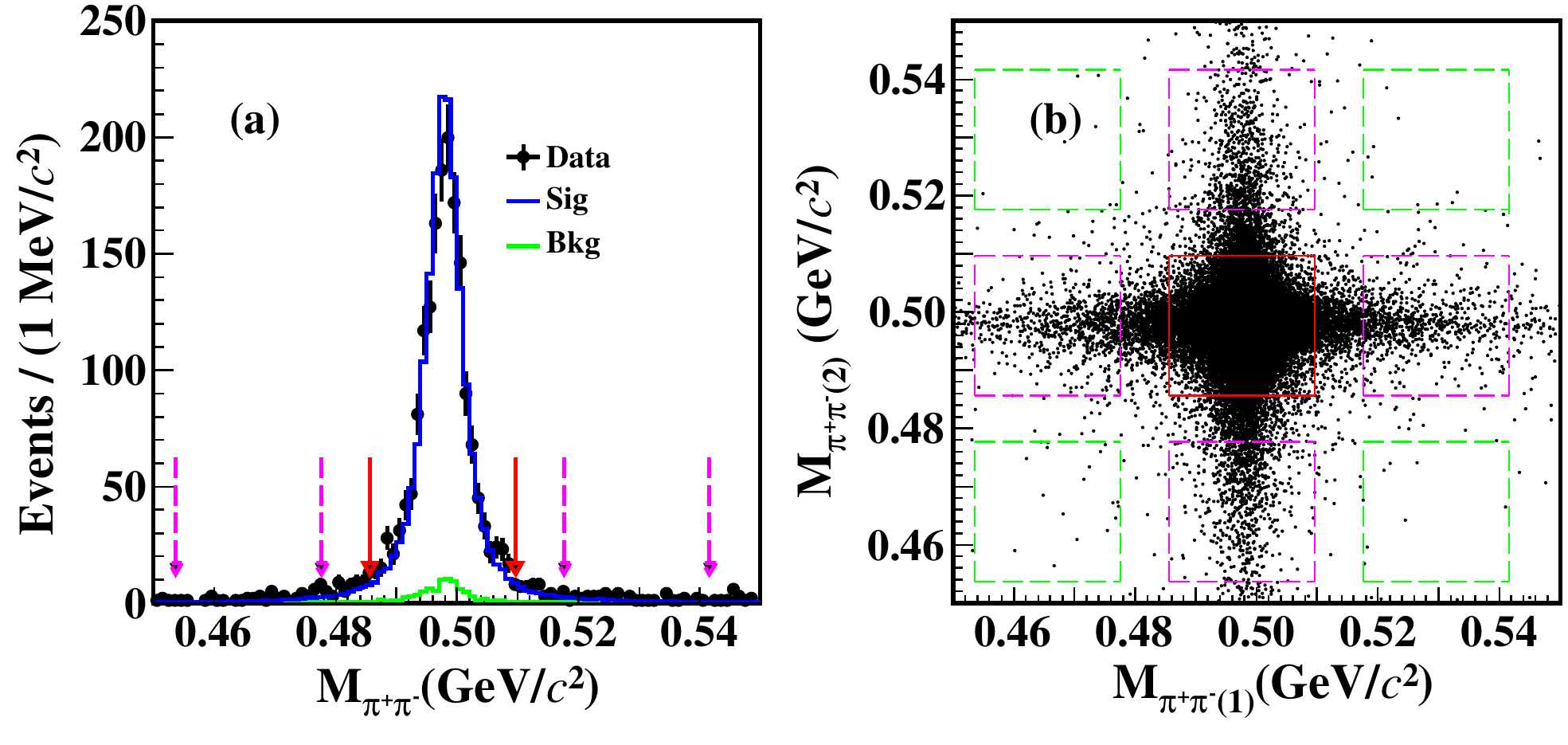}
\caption{
(a) The $M_{\pi^+\pi^-}$ distribution of the $K_S^0$ candidates in data and the scaled signal and background contributions from the inclusive MC sample,
where the pairs of red and pink arrows show the 1D $K_{S}^{0}$ signal and sideband regions, respectively.
(b) The $M_{\pi^+\pi^-(1)}$ versus $M_{\pi^+\pi^-(2)}$ distribution of the $K_S^0$ candidates in data, where
the red solid, pink dashed, and green long-dashed rectangles denote the 2D signal region, sideband 1 region, and sideband 2 region, respectively.
Besides all requirements mentioned in text, the events displayed here have been required to be within the region $(0.777,0.807)$~GeV/$c^2$ in the invariant mass of the $\pi^+\pi^-\pi^0$ system.
}
\label{tab:Sideband}
\end{figure}

The one-dimensional $K_{S}^{0}$ signal and sideband regions are defined as $(0.486,0.510)$ and $(0.454,0.478)$  or $(0.518,0.542)$ GeV/$c^2$, respectively.
The two-dimensional (2D) $K^0_SK^0_S$ signal region (called 2D signal region) is defined as the rectangle with both $\pi^+\pi^-$ combinations lying in the $K^0_S$ signal regions.
The 2D sideband regions (sideband 1 and sideband 2) are defined based on the one-dimensional sideband as shown in Fig.~\ref{tab:Sideband}(b).

\begin{figure*}[htbp]
\centering
\includegraphics[width=1.0\linewidth]{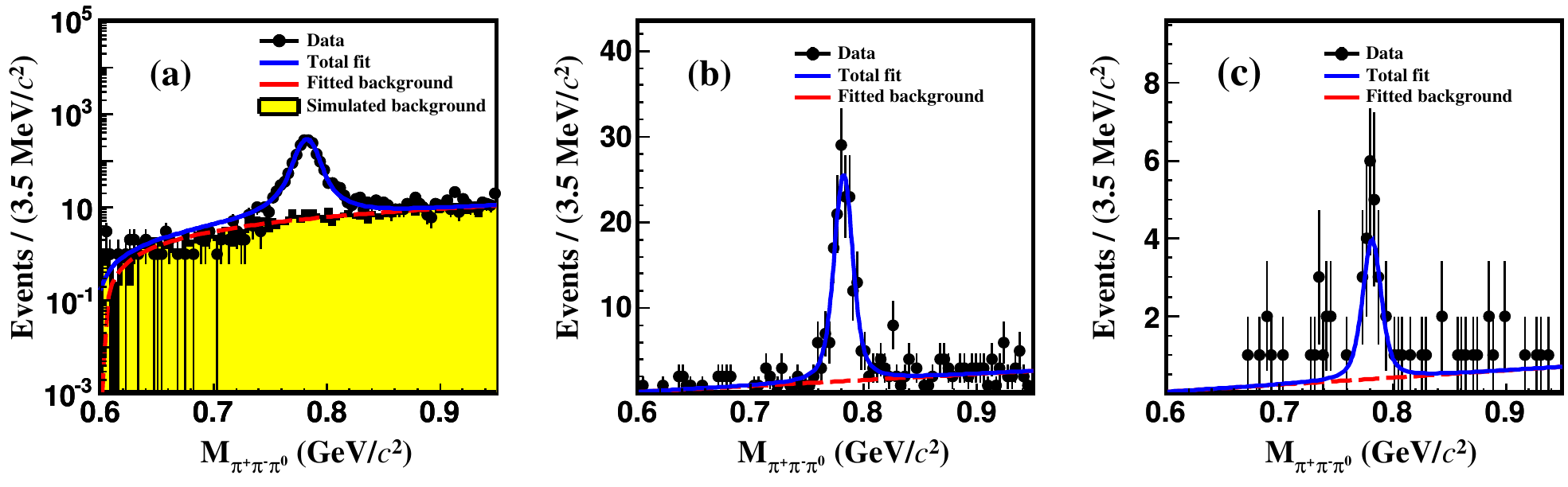}
\caption{Fits to the $M_{\pi^+\pi^-\pi^0}$ distributions of the accepted candidates in the 2D (a) signal, (b) sideband 1 and (c) sideband 2 regions for the $\psi(3686)$ data.
The black points with uncertainties are data.
The blue solid curves are the fit results.
The red dashed curves are the background components of the fit.
The yellow hatched histogram is the simulated background excluding those components estimated by the 2D sidebands.}
\label{tab:FitData}
\end{figure*}

Figure~\ref{tab:FitData} shows the $M_{\pi^+\pi^-\pi^0}$ distributions of the candidate events for  $e^+e^-\to \pi^+\pi^-\pi^0K_{S}^{0}K_{S}^{0}$ for the signal and sideband regions. In these figures, a clear $\omega$ signal is observed.
Unbinned maximum likelihood fits are performed to these events. In the fits, the $\omega$ signal shape is described by a Breit-Wigner function convolved with a Gaussian function and the background shape is described by a first order Chebychev polynomial. The width of the Breit-Wigner function is fixed to the world average value of the $\omega$ meson ~\cite{ref::pdg2020}. The mass of the $\omega$ meson as well as the mean and width of the Gaussian are free parameters. In the fits to the events in the 2D sideband 1 and sideband 2 regions, the parameters of the Gaussian function are fixed to the values obtained from the fit to the 2D signal region. The results of the best  fits to the  data are shown in Fig.~\ref{tab:FitData}.

From these fits, we obtain the numbers of $e^+e^-\to\omega K_S^0K_S^0$ in the 2D $K_S^0$ signal, sideband 1 and sideband 2 region, denoted as $N_{\rm sig}$, $N_{\rm side1}$, and $N_{\rm side2}$, respectively{\color{blue}, 
{\color{blue}where $N_{\rm side2}$ represents the yield of background events with two mis-formed $K_S^0$s, $N_{\rm side1}$ indicates that with one or two mis-formed $K_S^0$s, and $N_{\rm sig}$ is the number of events with two successfully formed $K_S^0$s. 
Assuming that the combinatorial backgrounds distribute evenly in the 2D spectrum and in the projected distribution within $K^0_S$ signal region,} the total number of $e^+e^-\to\omega K_S^0K_S^0$ candidate events in the $\psi(3686)$ data is calculate by 
%Thus,  if we want to get the number of $e^+e^-\to \omega K_S^0K_S^0$ candidate events ($N_{\rm tot}$) in the $\psi(3686)$ data, we have to subtract the events without $K_S^0$ ($N_{\rm side2}$) and those with only one $K_S^0$ ($N_{\rm side1}-N_{\rm side2}$) from $N_{\rm sig}$. 
%The background events with one or two mis-formed $K^0_S$ mesons distribute evenly along the $K^0_S$ signal region belt or in the whole plot, and they can be estimated by using the events locating in sideband 1 and sideband 2, respectively. Assuming that the combinatorial background is flat, the total number of $e^+e^-\to\omega K_S^0K_S^0$ candidate events in the $\psi(3686)$ data is calculate by 
\begin{equation}
\begin{split}
    %{N_{\rm tot}} ={N_{\rm sig}} - \frac{1}{2}({N_{side1}}-{N_{side2}}) -\frac{1}{4}{N_{side2}} ={N_{\rm sig}} - \frac {1}{2}{N_{\rm side1}} + \frac {1}{4}{N_{\rm side2}} .
     { N_{\rm tot}} &= { N_{\rm sig}} - \frac {1}{2}({ N_{\rm side1}} - { N_{\rm side2}}) - \frac {1}{4}{N_{\rm side2}}
 \\ &= {N_{\rm sig}} - \frac {1}{2}{N_{\rm side1}} + \frac {1}{4}{N_{\rm side2}},
\label{tab:Netnumber}
\end{split}
\end{equation}
The obtained values of $N_{\rm sig}$, $N_{\rm side1}$, $N_{\rm side2}$, and $N_{\rm tot}$ in data are summarized in the first row of Table~\ref{tab:Nnet}.}

\begin{figure*}[htbp]
\centering
\includegraphics[width=1.0\linewidth]{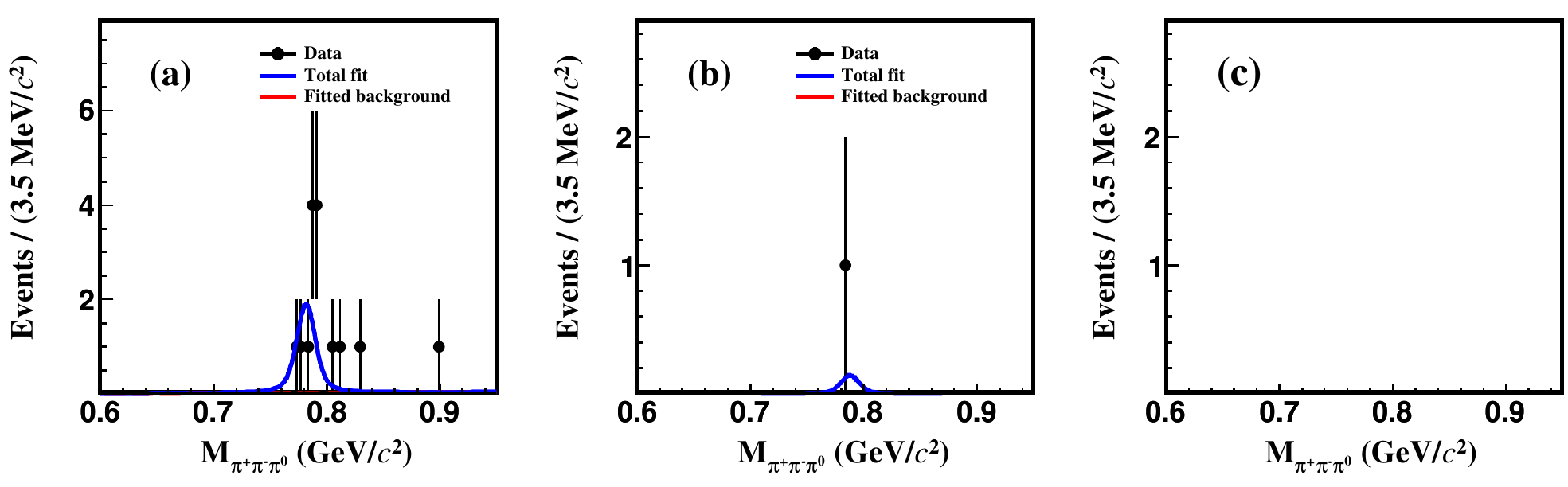}
\caption{Fits to the $M_{\pi^+\pi^-\pi^0}$ distributions  of the accepted candidates in the 2D (a) signal, (b) sideband 1 and (c) sideband 2 regions for the data at $\sqrt{s}=3.65 ~\mathrm{GeV}$.
The black points with uncertainties are data.
The blue solid curves are the fit results.
The red dashed curves are the background components of the fit.}
\label{tab:FitQED}
\end{figure*}

A similar analysis and fit are performed for the continuum data at $\sqrt{s}=3.65$ GeV.
The parameters of the Gaussian function are fixed to the values obtained from the fit to the $\psi(3686)$ data.
The results of the best fit to the data are shown in Fig. \ref{tab:FitQED}.
The obtained values of $N_{\rm sig}$, $N_{\rm side1}$, $N_{\rm side2}$, and $N_{\rm tot}$ in the continuum data are summarized in the second row of Table~\ref{tab:Nnet}.

\begin{table}[htbp]
   \centering
    \caption{The numbers of $N_{\rm sig}$, $N_{\rm side1}$, $N_{\rm side2}$, and $N_{\rm tot}$ obtained from $\psi(3686)$ data and continuum data, which were taken at $\sqrt{s} =$ 3.686 and 3.65 GeV, respectively. The uncertainties are statistical only.}
   \begin{tabular}{ccccc}
   \hline
   \hline
   $\rm E_{cm}$(GeV) & $N_{\rm sig}^{\rm fit}$ & $N_{\rm side1}^{\rm fit}$ & $N_{\rm side2}^{\rm fit}$ & $N_{\rm tot}$ \\
  \hline
3.686& $1811\pm46$   & $168\pm15$      & $25\pm6$        & $1733\pm47$   \\
3.650& $17.9\pm4.4$  & $1.0\pm1.2$     & 0.0           & $17.4\pm4.4$  \\
   \hline
   \hline
         \end{tabular}
     \label{tab:Nnet}
\end{table}

\begin{figure*}[htbp]
\centering
\includegraphics[width=0.9\linewidth]{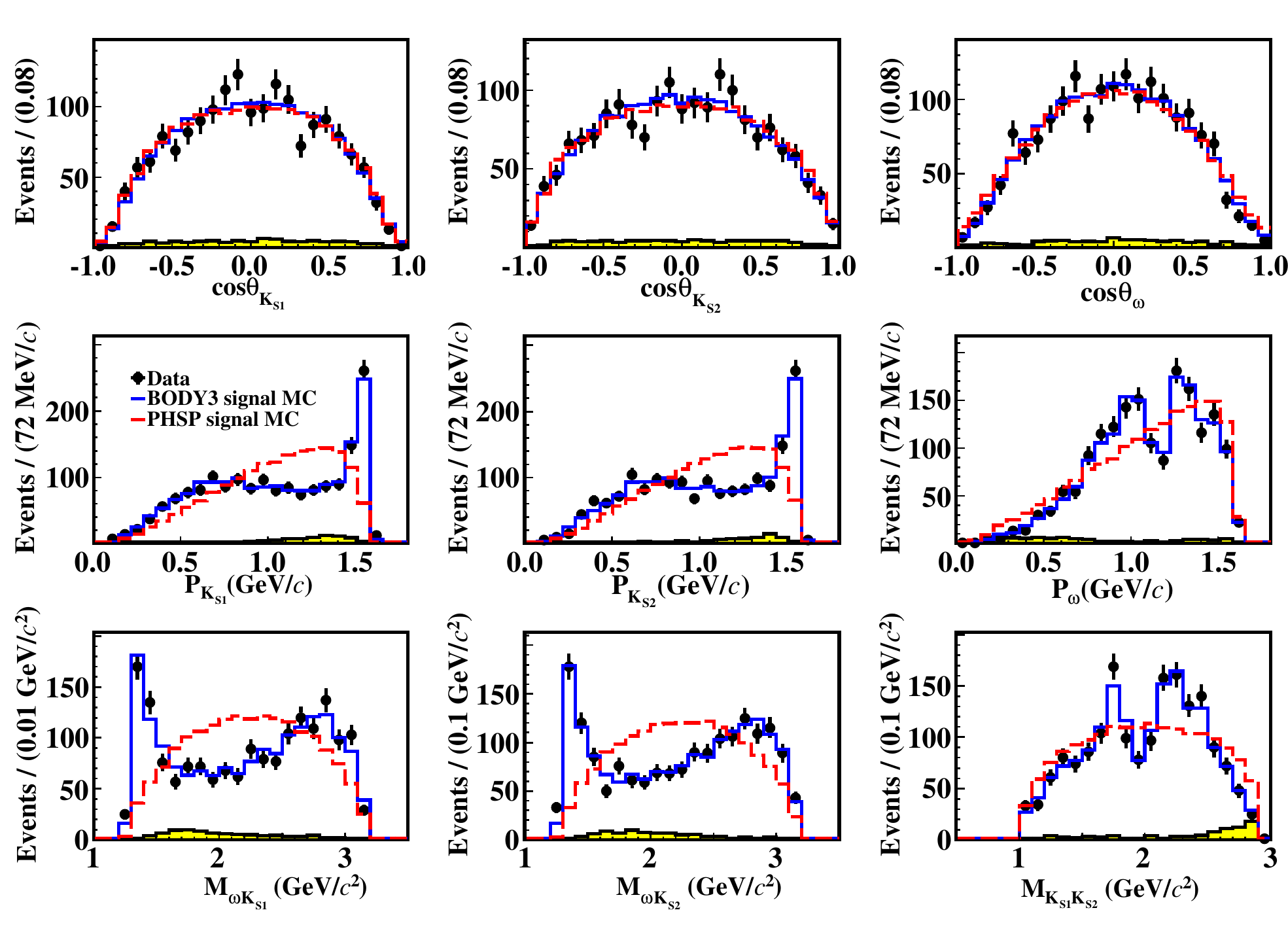}
\caption{
Distributions of $\rm cos\theta$, the momenta of daughter particles, and the invariant masses of all two-particle combinations of the $\psi(3686)\to \omega K^0_SK^0_S$ candidates. The points with uncertainties are data.
The blue solid curves and red dashed curves are BODY3 and PHSP signal MC events, respectively.
The yellow hatched histogram is the simulated background.
The signal and background yields have been normalized to data.
Besides all requirements mentioned in the text, the events shown here have been required to be within $(0.777,0.807)$~GeV/$c^2$ in the $\pi^+\pi^-\pi^0$ invariant mass.}
\label{tab:BODY3PHSP}
\end{figure*}

\begin{figure*}[htbp]
\centering
\includegraphics[width=0.9\linewidth]{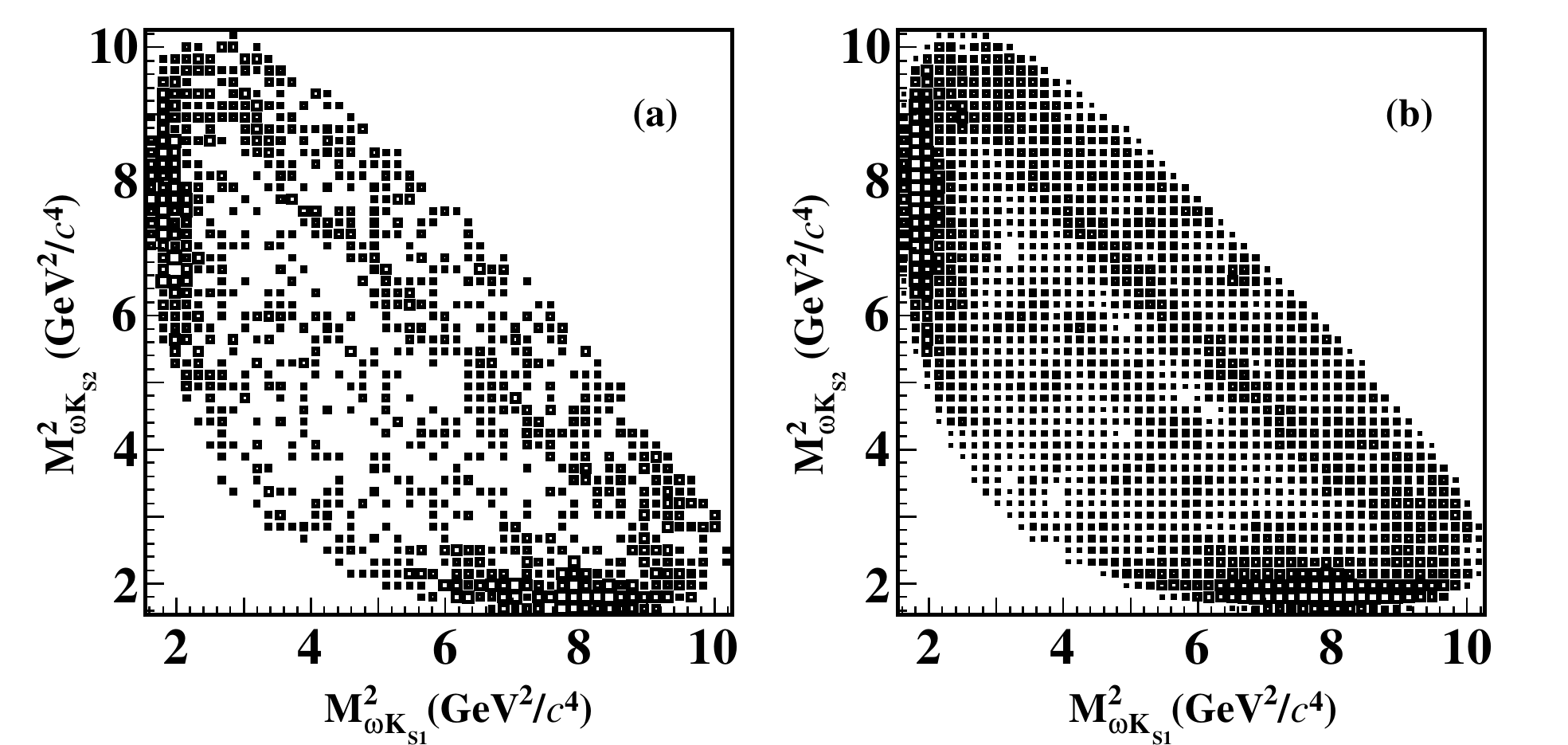}
\caption{Dalitz plots of $ M^2_{\omega  K_{S1}}$ vs $ M^2_{\omega  K_{S2}}$ of the accepted candidates in (a) data and (b) BODY3 signal MC events. 
}
\label{tab:dalitz}
\end{figure*}

Figure~\ref{tab:BODY3PHSP} shows the distributions of $\cos\theta$, where $\theta$ is the polar angle in the lab frame, and the momenta of final state particles as well as the invariant mass spectra of all two-particle combinations for those candidates surviving all selection criteria in data and MC simulation. Figure~\ref{tab:dalitz} shows the Dalitz plots of the accepted candiates in data and BODY3 signal MC events.
The $K_1(1270)$ resonance is observed in the $\omega K^0_S$ mass spectrum in the data.
To determine the efficiency ($\epsilon_{\psi(3686)\to\omega K_S^0K_S^0}$) for the signal process, the $\psi(3686)\to\omega K_S^0K_S^0$ decays are simulated with the modified data-driven generator BODY3~\cite{ref:evtgen}, which was developed to simulate different intermediate states in data for a given three-body final state. The Dalitz plot of $ M_{\omega K_S^0(1)}$ vs. $ M_{\omega K_S^0(2)}$ found in data, including a bin-wise correction for backgrounds and efficiencies, is taken as an input to the BODY3 generator.
The consistency between data and BODY3 signal MC events is much better than that between data and phase space (PHSP) signal MC events.
The BODY3 signal MC events are used to determine the signal selection efficiency in the further analysis.
{\color{blue}With the same analysis procedure as data analysis, the} detection efficiency for $\psi(3686)\to\omega K_{S}^{0}K_{S}^{0}$ is determined to be $(11.23\pm 0.05)\%$,
where the branching fractions of $\omega\to \pi^+\pi^-\pi^0$ and $K^0_S\to \pi^+\pi^-$ are not included. {\color{blue}The reliability of the detection efficiency is validated due to good consistency between data and MC simulation, as shown in Fig. 4.}

Under the assumption that there is no interference between the $\psi(3686)$ decay and the continuum production $e^+e^-\to \omega K_{S}^{0}K_{S}^{0}$, the branching fraction of $\psi(3686)\to \omega K_{S}^{0}K_{S}^{0}$ is determined by
\begin{widetext}
\begin{equation}
\mathcal{B}_{\psi(3686)\to\omega K_{S}^{0}K_{S}^{0}}=\frac{N^{\rm tot}_{\psi(3686)\to\omega K_{S}^{0}K_{S}^{0}}}{N_{\psi(3686)}\mathcal{B}_{\omega\to\pi^+\pi^-\pi^0} \mathcal{B}^{2}_{K_{S}^{0}\to \pi^{+}\pi^{-}}\mathcal{B}_{\pi^0\to\gamma\gamma} \epsilon_{\psi(3686)\to\omega K_{S}^{0}K_{S}^{0}}},
\end{equation}
\end{widetext}
where  $N^{\rm tot}_{\psi(3686)\to\omega K_{S}^{0}K_{S}^{0}}=N_{\rm tot}^{3.686}- N_{\rm tot}^{3.65}\times f_c$ is the total number of $\psi(3686)\to \omega K_{S}^{0}K_{S}^{0}$. It is determined to be $1495\pm81$.
 $f_c =\frac{{\mathcal L}_{3.686}}{{\mathcal L}_{3.65}}\times\frac{3.65^{2n}}{3.686^{2n}} \times\frac{\epsilon^{\rm QED}_{3.65}}{\epsilon^{\rm QED}_{3.686}} $ is a factor to take into account different luminosities~\cite{ref::psip-num-inc}, energies,  and detection efficiencies for the continuum production of $e^+e^-\to \omega K^0_SK^0_S$ at $\sqrt{s}=$ 3.686 and 3.65 GeV. Here, $n$ is a factor to consider $\frac{1}{s}$ dependence of cross section. We take $n=1$ and obtain $f_c = 13.7\pm0.03$.
$N_{\psi(3686)}$ is the total number of $\psi(3686)$ events.
$\mathcal{B}_{\omega\to\pi^+\pi^-\pi^0}$, $\mathcal{B}_{\pi^0\to\gamma\gamma}$ and $\mathcal{B}_{K_{S}^{0}\to \pi^{+}\pi^{-}}$ are the world average values of the branching fractions for $\omega\to\pi^+\pi^-\pi^0$, $\pi^0\to\gamma\gamma$ and $K_{S}^{0}\to \pi^{+}\pi^{-}$ from the PDG~\cite{ref::pdg2020}.
$\epsilon_{\psi(3686)\to\omega K_S^0K_S^0}$ is the detection efficiency for $\psi(3686)\to \omega K^0_SK^0_S$. We obtain ${\mathcal B}_{\psi(3686)\to \omega K^0_SK^0_S}=(7.04\pm0.39)\times 10^{-5}$.

\section{Systematic uncertainty}

In this section, the systematic uncertainties of the measurement of the $\psi(3686)\to\omega K_{S}^{0}K_{S}^{0}$ branching fraction are discussed.

The total number of $\psi(3686)$ events and its relative uncertainty are taken from Ref.~\cite{ref::psip-num-inc}. Its uncertainty of 0.65\% is considered as a systematic uncertainty. The statistical uncertainty is negligible.

The $\pi^\pm$ tracking efficiency was studied using a control sample of
$\psi(3686)\to\pi^+\pi^- J/\psi$ with $J/\psi\to \ell^{+}\ell^{-}$.
The difference in the tracking efficiencies between data and MC simulation is 1$\%$ per $\pi^\pm$.
The $\pi^\pm$ PID efficiency was investigated with a control sample of $J/\psi\to\rho\pi$ \cite{ref::pid}.
The difference in the PID efficiencies between data and MC simulation is 1\% per $\pi^\pm$.

The systematic uncertainty related to the photon selection efficiency is 1.0\% per photon,
which is determined using a control sample of $J/\psi\to\rho^0\pi^0$ with $\rho^0\to\pi^+\pi^-$ \cite{ref::pid}.

The systematic uncertainty originating from the $\pi^0$ mass window is examined by comparing the branching fractions measured with and without this requirement.
The change of the obtained branching fraction, 0.7\%, is taken as an uncertainty.

The $K^0_S$ reconstruction efficiency, including the efficiencies of tracking for $\pi^+\pi^-$, the $M_{\pi^+\pi^-}$ mass requirement, and the decay length requirement
was studied using control samples of $J/\psi\to$ $K^*(892)^\pm$ $K^{\pm}$, $K^*(892)^\pm$$\to K^0_S\pi^{\pm}$ and $J/\psi\to\phi K_{S}^{0}K^{\mp}\pi^{\pm}$~\cite{ref::ks_0}.
The difference in the $K^0_S$ reconstruction efficiencies between data and MC simulation is $1.6\%$ per $K^0_S$.
To estimate the systematic uncertainty related to the $K_S^0$ sideband, we examine the branching fraction using a $K_S^0$ sideband region shifted by $\pm4$ MeV/$c^2$. The maximum change of the branching fraction of 0.3\% is taken as the systematic uncertainty.

For the fit to the $M_{\pi^+\pi^-\pi^0}$ spectrum, three potential sources of systematic uncertainties are considered.
First, the uncertainty from the background shape is estimated using an alternative background shape determined from the inclusive MC sample. The change of the fitted signal yield, 0.1\%, is assigned as an uncertainty.
Second, the uncertainty from the signal shape is estimated by varying the fixed width of the $\omega$ meson by $\pm 1\sigma$, where $\sigma$ is the uncertainty of the world average value.
The change of the fitted signal yield is negligible.
Third, the uncertainty due to the fit range is estimated using fit ranges of $(0.55, 0.95)$, $(0.60, 1.00)$, $(0.65, 0.95)$, and $(0.65,0.90)$~GeV/$c^2$ instead.
The maximum change of the fitted signal yield, 0.8\%, is taken as an uncertainty.
Adding these three uncertainties in quadrature leads to the systematic uncertainty due to the fit to the $M_{\pi^+\pi^-\pi^0}$ spectrum of $0.8\%$.

The systematic uncertainty of the signal modeling with the BODY3 generator is estimated by varying the bin size of the input Dalitz plot by $\pm20$\%.
The maximum change of the signal efficiency, 0.8\%, is taken as the corresponding uncertainty. %{\color{blue}This is partially validated because the difference of the signal efficiencies between the BODY3 and PHSP models is only about 2.0\%.}

The systematic uncertainty due to the 5C kinematic fit is evaluated by varying the helix parameters by $\pm1$ standard deviations. The change of signal efficiency, 0.9\%, is taken as the relevant systematic uncertainty. 
The systematic uncertainty related to the scale factor $f_c$ for the continuum production of $e^+e^-\to \omega K^0_SK^0_S$ is estimated by comparing the difference between $1/s$ and $1/s^3$ dependencies of the cross section of the process $e^+e^-\to \omega K_S^0 K_S^0$. The effect on the branching fraction, $0.6\%$, is taken as the corresponding systematic uncertainty.

The uncertainties of the quoted branching fractions for $\omega\to\pi^+ \pi^-\pi^0$ and $K^0_S\to \pi^+\pi^-$ are 0.8\% and 0.1\%, respectively.
The uncertainty due to the limited MC statistics is considered as a source of systematic uncertainty.

  \begin{table}[htbp]
   \centering
       \caption{Systematic uncertainties of the branching fraction measurement.}

   \begin{tabular}{lc}
   \hline
   \hline
  Source  & Uncertainty (\%)\\
        \hline
  $N_{\psi(3686)}$                       & 0.65\\
  $\pi^\pm$ tracking                     & 2.0\\
  $\pi^\pm$ PID                          & 2.0\\
  $\pi^0$ reconstruction                 & 2.0\\
  $\pi^0$ mass window               & 0.7\\
  $K_{S}^{0}$ reconstruction             & 3.2\\
  $K_S^0$ sideband   &0.3\\
  $M_{\pi^+\pi^-\pi^0}$ fit                         & 0.8\\
  BODY3 generator                        & 0.8\\
  5C kinematic fit                       & 0.9\\
  $f_{c}$ factor                         & 0.6\\
$\mathcal{B}_{\omega\to\pi^+\pi^-\pi^0}$ & 0.8\\
$\mathcal{B}_{K_{S}^{0}\to\pi^+\pi^-}$   & 0.2\\
  MC statistics                          & 0.4\\
   Total & 5.2\\
   \hline
   \hline
         \end{tabular}
     \label{tab:Sys}
\end{table}

Assuming that all sources are independent, the total systematic uncertainty is determined to be 5.2\%.
The systematic uncertainties are summarized in Table \ref{tab:Sys}.

\section{Summary}

Using $(448.1\pm2.9)\times10^6$ $\psi(3686)$ events accumulated with the BESIII detector, we measure the branching fraction of $\psi(3686)\to\omega K_{S}^{0}K_{S}^{0}$ for the first time. We find
\begin{equation}
\mathcal{B}_{\psi(3686)\to\omega K_{S}^{0}K_{S}^{0} }= (7.04 \pm 0.39 \pm 0.37)\times10^{-5},
\nonumber
\end{equation}
where the first  uncertainties are statistical and the second are systematic.
This branching fraction is determined under the assumption that interference between
$\psi(3686)\to\omega K_S^0 K_S^0$ and the continuum process $e^+e^-\to\omega K_S^0 K_S^0$
is negligible.
Combining with the world average value of $\mathcal{B}_{J/\psi\to\omega K\bar K}=(1.90\pm0.40)\times 10^{-3}$~\cite{ref::pdg2020} and taking isospin symmetry into account, we obtain the ratio
\begin{equation}
\mathcal{Q}_{\omega K_{S}^{0}K_{S}^{0}}=\frac{\mathcal{B}_{\psi(3686)\to\omega K_{S}^{0}K_{S}^{0}}}{\mathcal{B}_{J/\psi\to\omega K_{S}^{0}K_{S}^{0} }} = (14.8 \pm 3.2)\%.
\nonumber
\end{equation}
This value is consistent with the 12\% rule within the uncertainty. 
Combining with the branching ratio $\mathcal{B}_{\psi(3686)\to\omega K^{+}K^{-}}=(15.6 \pm 0.4 \pm 1.1) \times10^{-5}$ from Ref.~\cite{bes3_omegaKK}, we determine
\begin{equation}
\frac{\mathcal{B}_{\psi(3686)\to\omega K^{+}K^{-}}}{ 2\cdot \mathcal{B}_{\psi(3686)\to\omega K_S^{0}K_S^{0}}} = 1.11 \pm 0.07\pm0.07,
\nonumber
\end{equation}
which indicates that a possible violation of isospin symmetry
in this decay is too small to be observed at present.
Amplitude analyses of $\psi(3686)\to\omega K^+K^-$ and  $\psi(3686)\to\omega K_{S}^{0}K_{S}^{0}$
with larger data samples in the near future
will help to explore the intermdiate states in these decays and thereby reveal if there is possible isospin symmetry violation in specific local kinematic regions.

\section{Acknowledgement}
The BESIII collaboration thanks the staff of BEPCII and the IHEP computing center for their strong support. This work is supported in part by National Key Research and Development Program of China under Contracts Nos. 2020YFA0406300, 2020YFA0406400; National Natural Science Foundation of China (NSFC) under Contracts Nos. 11875170, 11775230, 11625523, 11635010, 11735014, 11822506, 11835012, 11935015, 11935016, 11935018, 11961141012, 12022510, 12025502, 12035009, 12035013, 12061131003;  the Chinese Academy of Sciences (CAS) Large-Scale Scientific Facility Program; Joint Large-Scale Scientific Facility Funds of the NSFC and CAS under Contracts Nos. U1832207, U1732263; CAS Key Research Program of Frontier Sciences under Contract No. QYZDJ-SSW-SLH040; 100 Talents Program of CAS; INPAC and Shanghai Key Laboratory for Particle Physics and Cosmology; ERC under Contract No. 758462; European Union Horizon 2020 research and innovation programme under Contract No. Marie Sklodowska-Curie grant agreement No 894790; German Research Foundation DFG under Contracts Nos. 443159800, Collaborative Research Center CRC 1044, FOR 2359, FOR 2359, GRK 214; Istituto Nazionale di Fisica Nucleare, Italy; Ministry of Development of Turkey under Contract No. DPT2006K-120470; National Science and Technology fund; Olle Engkvist Foundation under Contract No. 200-0605; STFC (United Kingdom); The Knut and Alice Wallenberg Foundation (Sweden) under Contract No. 2016.0157; The Royal Society, UK under Contracts Nos. DH140054, DH160214; The Swedish Research Council; U. S. Department of Energy under Contracts Nos. DE-FG02-05ER41374, DE-SC-0012069.

\end{document}